\documentclass[prl,reprint,twocolumn,showpcs,amsmath,amssymb,superscriptaddress]{revtex4}
\usepackage{graphicx}
\usepackage{dcolumn}
\usepackage{bm}

\begin{document}
\title{Full Electrical Control of the Electron Spin Relaxation in GaAs Quantum Wells} 
\author{A. Balocchi}
\author{Q. H. Duong}
\author{P. Renucci}
\affiliation{Universit\'e de Toulouse, INSA-CNRS-UPS, LPCNO, 135 avenue de Rangueil, 31077 Toulouse, France}
\author{B. Liu}
\affiliation{Beijing National Laboratory for Condensed Matter Physics, Institute of Physics, Chinese Academy of Sciences, P.O. Box 603, Beijing 100190, PR China}
\author{C. Fontaine}
\affiliation{LAAS, CNRS, Universit\'e de Toulouse; 7 avenue du Colonel Roche, F-31077 Toulouse, France }
\author{T. Amand}
\author{D. Lagarde}
\author{X. Marie}
\email{marie@insa-toulouse.fr}
\affiliation{Universit\'e de Toulouse, INSA-CNRS-UPS, LPCNO, 135 avenue de Rangueil, 31077 Toulouse, France}
\date{\today}
\begin{abstract}
The electron spin dynamics in (111)-oriented GaAs/AlGaAs quantum wells is studied by time-resolved photoluminescence spectroscopy. By applying an external field of 50 kV/cm a two-order of magnitude increase of the spin relaxation time can be observed reaching values larger than 30 ns; this is a consequence of the electric field tuning of the spin-orbit conduction band splitting which can almost vanish when the Rashba term compensates exactly the Dresselhaus one. The measurements under transverse magnetic field demonstrate that the electron spin relaxation time for the three space directions can be tuned simultaneously with the applied electric field.
\end{abstract}
\maketitle

The control of the electron spins in semiconductors for potential use in transport devices or quantum information applications has attracted a great attention in recent years~\cite{Meier1984,awschalom_semiconductor_2010,Dyakonov_book}. In 2D nanostructures made of III-V or II-VI semiconductors, the dominant loss of electron spin memory is related to the spin relaxation mechanism known under the name D’yakonov-Perel (DP)~\cite{Dyakonov1972,Dyakonov1986}. In these materials, the absence of inversion symmetry and the spin-orbit (SO) coupling are responsible for the lifting of degeneracy for spin $\mid\! \! 1/2\rangle$ and  $\mid\! \! -1/2\rangle$ electrons states in the conduction band (CB). This splitting  plays a crucial role for the spin manipulation and spin transport phenomena~\cite{zutic,ivchenko_optical_2005}. As it depends strongly on the crystal and nanostructure symmetry~\cite{dresselhaus_spin-orbit_1955,bychkov_oscillatory_1984,krebs_giant_1996}, it can be efficiently tailored as explained below.
The SO splitting can be viewed as the result of the action on the electron spin of an effective magnetic field whose amplitude and direction depend on the wave vector $\boldsymbol{k}$ of the electron. The electronic spin will precess around this field with an effective, momentum dependent, Larmor vector $\boldsymbol{\Omega}$ whose magnitude corresponds to the CB spin splitting.  This effective magnetic field changes with time since the direction of electron momentum varies due to electron collisions. As a consequence, spin precession around this field in the intervals between collisions gives rise to spin relaxation. In the usual case of frequent collisions, the relaxation time of an electron spin oriented along the direction $i$ can be written~\cite{Dyakonov1972}: $(\tau_s^i)^{-1}=\langle \Omega_{\perp}^2\rangle\tau_p^{*}$,
where $\langle \Omega_{\perp}^2\rangle$ is the mean square precession vector in the plane perpendicular to the direction $i$  ($i$=$x,y,z$) and $\tau_p^{*}$ the electron momentum relaxation time. This yields the loss of the electron spin memory in a few tens or hundreds of picoseconds~\cite{malinowski_spin_2000,amand_spin_1997}. As the driving force in the DP spin relaxation is the SO splitting, its reduction is expected to lead to an increase of the spin relaxation time~\cite{bu_long_2010,Lagarde2008}.
In bulk zinc blende semiconductor, the Bulk Inversion Asymmetry (BIA) spin splitting, also called Dresselhaus term, is determined by~\cite{Meier1984,dresselhaus_spin-orbit_1955}:
\begin{small}
\begin{equation}
\boldsymbol{\Omega}_{BIA}^{3D}\left(\boldsymbol{k}\right)=\frac{\gamma}{\hbar}\left[k_x \left(k_y^2-k_z^2\right), k_y\left(k_z^2-k_x^2\right), k_z\left(k_x^2-k_y^2\right)\right]\!,
\end{equation}
\end{small}
where $\gamma$ is the Dresselhaus coefficient and $\boldsymbol{k}\!=\!(k_x,k_y,k_z)$ the electron wavector.
In a quantum well (QW) where the momentum component along the growth axis $\mathbf{z}$ is quantized, the vector $\boldsymbol{\Omega}$ due to the BIA for the lowest electron sub-band writes:
\begin{subequations}
\begin{align}
\boldsymbol{\Omega}_{BIA}&\left(\boldsymbol{k_{\parallel}}\right)=\frac{\gamma}{\hbar}\langle k_z^2\rangle \left(-k_x,k_y,0\right)&\mathrm{if~~\mathbf{z} \parallel [001]}\\
\boldsymbol{\Omega}_{BIA}&\left(\boldsymbol{k_{\parallel}}\right)=\frac{\gamma}{2\hbar}\langle k_z^2\rangle \left(0,0,k_y\right)&\mathrm{if~~\mathbf{z} \parallel [110]}\\
\boldsymbol{\Omega}_{BIA}&\left(\boldsymbol{k_{\parallel}}\right)=\frac{2\gamma}{\hbar\sqrt{3}}\langle k_z^2\rangle \left(k_y,-k_x,0\right)&\mathrm{if~~\mathbf{z} \parallel [111]}
\end{align}
\end{subequations}
where $\langle k_z^2\rangle$  is the averaged squared wavevector along the growth direction and $\boldsymbol{k_{\parallel}}$ the in-plane wavevector. The weaker cubic terms in the Hamiltonian have been neglected here since $\langle k_z^2\rangle  > k_x^2,k_y^2$; their influence  will be discussed later in the paper.
If an external electric field is applied or an asymmetric doping which generates a built-in electric field is present in the structure, an additional contribution to the spin splitting appears, usually referred to as Structural Inversion Asymmetry (SIA) or Rashba effect~\cite{bychkov_oscillatory_1984,averkiev_giant_1999,kainz_anisotropic_2003,miller_gate-controlled_2003,eldridge_spin-orbit_2011}.
For an electric field applied along the growth axis, this SIA term writes:
\begin{equation}
\boldsymbol{\Omega}_{SIA}\left(k_{\parallel}\right) = \frac{\alpha}{\hbar}\left(k_y,-k_x,0\right),
\end{equation}
where $\alpha= a\cdot E$ is the Rashba coefficicient ($E$ is the electric field amplitude and $a$ is a positive constant). Note that the SIA vector is perpendicular to both the electron momentum and  the electric field ($\boldsymbol{\Omega} \propto \boldsymbol{E} \times\boldsymbol{k}$).
As shown in equations (2) and (3) both the BIA and SIA strengths can be controlled~\cite{averkiev_giant_1999}. The modification of the electron quantum confinement (through the well width $L$ variation of the   $\langle k_z^2\rangle$ term for instance~\cite{malinowski_spin_2000}) allows the tuning of the BIA spin splitting whereas the SIA term can be easily varied through the applied electric field amplitude.

For (001) quantum wells, the interplay of the BIA and SIA terms leads to an in-plane anisotropic CB spin-splitting (eq. 2a and 3)~\cite{averkiev_giant_1999,jusserand_spin_1995,schliemann_nonballistic_2003,lechner_tuning_2009}. By tuning the QW asymmetry, longer DP spin relaxation times along the in-plane [110] direction was demonstrated~\cite{stich_detection_2007,averkiev_spin-relaxation_2006,liu_electron_2007,koralek_emergence_2009}.\\
For (110) quantum wells, it turns out that the effective magnetic field due to the BIA term is oriented along the [110] growth direction (equation 2b)~\cite{Dyakonov1986}; if the electron spin
 S$_z$ is also aligned along this direction, the DP spin relaxation mechanism is suppressed, leading to $\tau_s^z$, as long as 20 ns for electrons spins parallel to [110]~\cite{ohno_spin_1999,doehrmann_anomalous_2004}. This is not true for spins prepared in other directions due to strongly anisotropic spin relaxation in these (110)
   GaAs quantum wells~\cite{lau_tunability_2002,doehrmann_anomalous_2004,karimov_high_2003}.\\
In contrast to (100) and (110) QW, where the DP spin relaxation vanishes for one given spin direction, it can be suppressed in principle for the three directions in space for (111) quantum wells, since the conduction bands can become spin degenerate to first order in $\boldsymbol{k}$~\cite{cartoixa_suppression_2005,vurgaftman_spin-relaxation_2005,sun_spin_2010}. This is possible since the BIA and SIA vectors have exactly the same direction in the QW plane and the same $\boldsymbol{k}$ dependence (equation 2c and 3, inset of figure 2a). If the electric field amplitude $E$ is tuned to the value:
\begin{equation}
\frac{2\gamma}{\hbar\sqrt{3}}\langle k_z^2\rangle=-aE,
\end{equation}
the DP spin relaxation should be suppressed for $S_x$, $S_y$, $S_z$ since all components of $\boldsymbol{\Omega}$ vanish. To the best of our knowledge this has not been evidenced experimentally.

In this letter we have measured the electron spin relaxation time in (111) GaAs quantum wells by time-resolved photoluminescence (PL). By embedding the QWs in a PIN or NIP structure we demonstrate the tuning of the conduction band spin splitting and hence the spin relaxation time with an applied external electric field. The spin quantum beats measurements under transverse magnetic field prove that the DP spin relaxation time is not only increased for the $S_z$ spin component but also for both $S_x$ and $S_y$. 

\begin{figure}
 \includegraphics[width=0.45\textwidth,keepaspectratio=true]{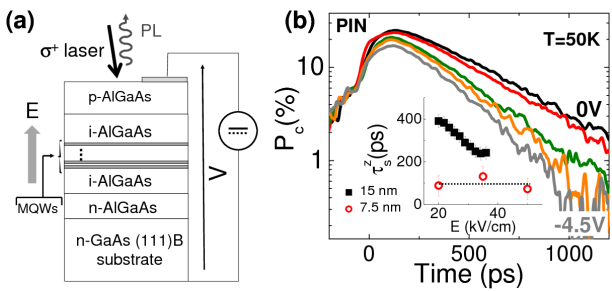}
 \caption{(Color online)  (a) PIN GaAs/AlGaAs MQW structure. $E$ is positive when it points along the growth direction. (b) Circular polarization dynamics for different reverse bias voltages $V$ for sample I ($L$=15 nm). inset: dependence of the electron spin relaxation time with the electric field for sample I and III ($L$=7.5 nm).} 
 \label{figure_1}
\end{figure}
We have studied three samples grown on n-doped (111)B GaAs substrates (n=2$\cdot$10$^{18}$cm$^{-3}$) by molecular beam epitaxy. These substrates are misoriented 3$^{\circ}$ towards $\langle$111$\rangle$A. 
For the three samples, the Multiple Quantum Well (MQW) region consists in 20 GaAs/Al$_{0.3}$Ga$_{0.7}$As QWs with a barrier width $L_B$=12 nm. Sample I and II have a well width equal to $L$=15 nm whereas $L$=7.5 nm for sample III. 
In samples I and III, these layers form a PIN device with the following structure sequence (figure 1a): n-substrate /500 nm GaAs layer (n=2$\cdot$10$^{18}$cm$^{-3}$)/500nm AlGaAs (n=2$\cdot$10$^{18}$cm$^{-3}$)/100 nm non intentionally doped (nid)  AlGaAs/MQW/100 nm nid AlGaAs/500 nm p  AlGaAs (p=2$\cdot$10$^{18}$cm$^{-3}$)/5nm p+ GaAs (p=1$\cdot$10$^{19}$cm$^{-3}$). The bias is applied between the surface p-contact and the substrate back n-contact.
In sample II, the n and p-doped layers are reversed in the sequence and a 500 nm nid GaAs layer followed by a 100 nm nid AlGaAs layer are grown in addition just above the substrate. In this NIP structure, the bias is applied between the un-etched top n-contact and the chemically etched p-contact. The built-in electric field $\mid\!E\!\mid$ in the PIN or NIP structure is about 20 kV/cm.
Applying a reverse bias in these samples allows us to tune the sign and amplitude of the Rashba (SIA) spin splitting. The electric field in the MQW region is determined from measurements of quantum confined Stark effect~\cite{tober_determining_1993,vinattieri_electric_1993}. The samples are excited by  circularly polarized ($\sigma^+$) 
1.5 ps pulses generated by a mode-locked Ti-Sa laser with a repetition frequency of 80 MHz (average power P$_{exc}$=10 mW, 50 $\mu$m diameter spot). The laser excitation energy is set to 1.56 eV, $i.e.$ above the E1-LH1 quantum well transition. The E1-HH1 transition time-resolved PL is then recorded using a S1 photocathode streak camera with an overall time resolution of 8 ps. The PL intensity of the E1-HH1 transition co-polarized ($I^+$) and counter-polarized ($I^-$) with the excitation laser is then recorded and the PL circular polarization degree is defined as $P_c = (I^+-I^-)/ (I^++I^-)$. 

Figure 1b displays the PL circular polarization dynamics $P_c $ for different bias voltages $V$ in sample I at T=50 K. Let us recall that $P_c$ probes directly the QW electron spin dynamics in these non-resonant excitation conditions~\cite{Meier1984,amand_spin_1997} since (i) the hole spin relaxation time is very fast (of the order of $\tau_h\sim$1 ps) and (ii) the exciton spin relaxation time due to electron-hole exchange interaction is inefficient due to the short $\tau_h$.  We observe in figure 1b a clear variation of the electron spin relaxation time with $V$: the relaxation time $decreases$ when the reverse bias increases, $i.e.$ when the electrons in the quantum wells feel a larger  electric field $E$. This variation rules out a possible role played by the exciton spin relaxation mechanism (induced by the exchange interaction). Indeed it was shown that the application of an external electric field when excitons are excited strictly resonantly yields an $increase$  of the polarization decay time due to the reduction of the exchange interaction (weaker overlap between the electron and hole wavefunction)~\cite{vinattieri_electric_1993}. Thus, the results in figure 1b demonstrate that the electron spin relaxation time in (111) quantum wells can be tuned by the Rashba effect. In the narrow MQW ($L$=7.5 nm, sample III), we see in the inset of figure 1b that the electric field has almost no effect on the spin relaxation compared to the one in sample I as a consequence of the larger BIA splitting due to the stronger $\langle k_z^2 \rangle=\left(\pi/L\right)^2$ term. However, in these PIN devices, it turns out that the Dresselhaus and Rashba $\boldsymbol{\Omega}$ vectors have the same direction and same sign. In other words, these two contributions to the conduction band spin splitting add up: $\boldsymbol{\Omega}$ increases with $E$ and the DP spin relaxation time decreases with the reverse bias.

In order to check this assumption, we have measured the electron spin dynamics in sample II where the applied electric field direction is reversed with respect to the one in
 sample I. In this case we expect to reduce the CB spin splitting (the BIA and SIA terms will subtract to each other) and as a consequence an $increase$ of the electron spin
  relaxation time with the applied reverse bias should be observed. Figure 2a presents the variation of the electron spin dynamics with the applied reverse bias in sample II (NIP
   structure). A spectacular increase of the electron spin relaxation time is observed when the electric field modulus $\mid\!E\!\mid$ increases. The  dependence of the electron spin relaxation time $\tau_s^z$ as a function of $E$ is displayed in figure 2b (circles): the electron spin relaxation time increases from 500 ps to $\sim$30 000 ps when $\mid\!E\!\mid$ 
     varies from 20 to 55 kV/cm. For $\mid\!E\!\mid>$ 50 kV/cm, the electron spin relaxation time ($\tau_s^z >$ 10 ns) becomes much longer than the carrier lifetime so that it is difficult to measure it in time-resolved PL. For larger $E$ we expect that an efficient DP spin relaxation should be switched on again because
the spin splitting would not be anymore zero~\cite{cartoixa_suppression_2005}. The NIP device characteristics did not allow us to test this prediction (55 kV/cm being the maximum achievable electric field).
Assuming that E= -55 kV/cm is close to the cancellation of the spin splitting on the basis of the very long measured $\tau_s^z$, we can extract from equation (4) the
ratio between the Dresselhaus $\gamma$ and the Rashba $\alpha$ coefficients. We find $\alpha/\gamma$= 5$\cdot$10$^{16}$ m$^{-2}$. No measurements were
made before in (111) QW for comparison. Nevertheless, using the measured $\gamma$ value in a (001) QW with similar confinement energy 
 ($\mid\!\gamma\!\mid$=11 eV$\cdot \mathrm{\AA}^3$)~\cite{leyland_oscillatory_2007},  gives a Rasha coefficient $\alpha$= 550$\cdot$10$^{-15}$ eV$\cdot$m, in good agreement with the coefficient measured in a (110) MQW sample with an applied field of 60 kV/cm ($\alpha$=500$\cdot$10$^{-15}$ eV$\cdot$m)~\cite{eldridge_all-optical_2008}. 
These results demonstrate that the electron spin relaxation time in (111) quantum wells can be tuned with an applied electric field thanks to the Rashba-induced variation of the CB spin splitting.

Nevertheless we have shown so far that only very long electron relaxation times for spins which are oriented along the [111] growth direction can be obtained with the reverse bias in the NIP structure. We recall that in the optical spin orientation experiments presented here, the photogenerated electron spin is parallel to the growth direction which is also the propagation direction of the excitation laser beam (inset of figure 2a).
\begin{figure}
 \includegraphics[width=0.45\textwidth,keepaspectratio=true]{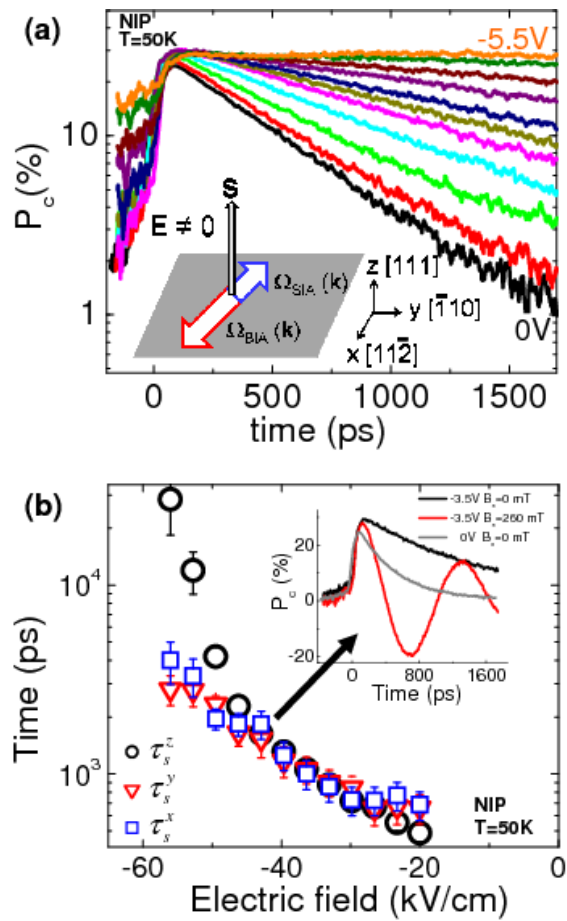}
 \caption{(Color online) Sample II (NIP) (a) Electron spin dynamics at T=50 K for different reverse voltages. Inset: Schematic of the BIA ($\boldsymbol{\Omega_{BIA}})$  and SIA ($\boldsymbol{\Omega_{SIA}}$)  precession vector in  (111) GaAs quantum wells. $\boldsymbol{S}$ is the photogenerated electron spin at $t$=0. (b) Dependence of the electron spin relaxation times $\tau_s^z$ (circles), $\tau_s^y$ (triangles),  $\tau_s^x$ (squares), as a function of the electric field. Inset: PL circular polarization dynamics for $B_x$=0 and $B_x$=0.26 T ($V$=-3.5V, see arrow). The $P_c$ dynamics for $V$=0 and $B_x$=0 is also displayed.} 
 \label{figure_2}
\end{figure}
In order to test that the relaxation time can also be controlled electrically for spins along the $\mathbf{x}$ and $\mathbf{y}$ in-plane QW direction, which is the key feature of (111) quantum wells, we have measured the electron spin dynamics in transverse magnetic field (Voigt configuration). Following the circularly polarized laser excitation, the electron spins will be first oriented along the $\mathbf{z}$ axis but will then precess around the in-plane external magnetic field. As a consequence, the electron spins should experience both in-plane and out-of plane spin relaxation processes. The inset of figure 2b depicts for illustration the PL circular polarization dynamics for $B_x$=0 and $B_x$=0.26 T in the NIP sample. In both cases, the same reverse bias $V$=- 3.5 V is applied (the decay for $V$=0 V is also displayed as a reference). For $B_x$= 0, the average electron spin direction is always oriented along the $\mathbf{z}$ direction; we measure $\tau_s^z$=1630 ps (compared to $\tau_s^z$=490 ps at $V$=0 V). For $B_x$=0.26 T, we observe the expected oscillations as a result of the electron spin Larmor precession around the external magnetic field $B_x$. The Larmor precession period yields the measured electron spin Land\'e factor $\mid \!g\! \mid$ = 0.22~\cite{note}. The striking feature here is that the circular polarization decay time in the presence of the external transverse magnetic field is almost the same as the one for $B_x$=0. Similar results have been obtained for different $B_x$ in the range 0.26-0.7 T. We measure a spin relaxation time  $\left(1/\tau_s^z + 1/\tau_s^y\right) /2$ which yields $\tau_s^y$=1530$\pm$300 ps. We have performed the same experiment for an applied magnetic field along the  $\mathbf{y}$ direction, which gives $\tau_s^x$= 1830$\pm$300 ps.

The large increase of $\tau_s^x$ and $\tau_s^y$ observed as a function of $\mid \!E\!\mid$ (see figure 2b) demonstrates that the DP spin relaxation time in (111) quantum wells can not only be controlled electrically for the $S_z$ electron spin component but also for both $S_x$ and $S_y$. These results contrast drastically with the (001) and (110) quantum wells~\cite{doehrmann_anomalous_2004,karimov_high_2003,stich_detection_2007,averkiev_spin-relaxation_2006,liu_electron_2007}. \\
Nevertheless we observe in figure 2b that the in-plane spin relaxation times are shorter than the out-of plane ones for large applied electric field ($\mid\! E \!\mid \geq$ 50 kV/cm). When both Rashba and Dresselhaus terms have similar amplitudes at a non-zero temperature, we believe that the main reason for a large spin relaxation anisotropy is due to the cubic Dresselhaus term which has been neglected in equation (3c). Taking into account this non-linear term, the total  CB spin splitting is determined by~\cite{cartoixa_suppression_2005,vurgaftman_spin-relaxation_2005,sun_spin_2010}:    
\begin{equation*}
\boldsymbol{\Omega}=\boldsymbol{\Omega}_{BIA}+\boldsymbol{\Omega}_{SIA}\equiv
\end{equation*}
\vskip -0.7cm
\begin{equation}
\left\{
\begin{array}{l}
\Omega_x \left(k\right)=\frac{1}{\hbar}\left(\frac{\gamma}{\sqrt{3}}\left(2\langle k_z^2\rangle - \frac{k^2}{2} \right) +aE \right) k_y  \\
\Omega_y \left(k\right)=-\frac{1}{\hbar}\left(\frac{\gamma}{\sqrt{3}}\left(2\langle k_z^2\rangle - \frac{k^2}{2}\right)+aE \right) k_x \\
\Omega_z \left(k\right)=\frac{1}{\hbar}\left(\frac{3k_x^2 k_y -k_y^3}{\sqrt{6}}\right)
\end{array}
\right.
\end{equation}
This equation shows that for a given $\boldsymbol{k}$ the strict compensation of Rashba and Dresselhaus term can only occur for the in-plane components $\Omega_x$ and $\Omega_y$. As a consequence, electrons spin oriented along the  $\mathbf{z}$ axis will have longer spin relaxation times compared to the ones oriented in the plane since the latter will precess around a non-zero vector  $\boldsymbol{\Omega$} which is tilted with respect to the QW plane~\cite{sun_spin_2010}. Moreover equation (5) demonstrates that the compensation can not strictly occur simultaneously for a large range of electron wavevectors. When the temperature increases, higher $k$ states are populated where the cubic Dresselhaus term plays a larger role~\cite{vurgaftman_spin-relaxation_2005}.
Despite this non-linear Dresselhaus term we observed a clear control of the SO spin splitting in the 15 nm MQW at high temperature. A factor of five increase of the spin relaxation time $\tau_s^z$ is still observed when the bias varies from 0 to -3 V at T=150 K (close to the ratio observed at T=50 K). Above 150 K, it was not possible to separate the  Stark-shifted QW and the bulk GaAs luminescence, preventing us to get room temperature data. For larger quantum wells at high temperature one also has to consider other spin relaxation mechanisms due to the scattering of electrons between different QW subbands~\cite{doehrmann_anomalous_2004,bernardes_spin-orbit_2007}.\\
We emphasize that the tuning or suppression of the DP electron spin relaxation demonstrated here for GaAs/AlGaAs quantum wells is also possible in many other III-V and II-VI zinc-blende nanostructures since the principle relies only on symmetry considerations. Spin transport of electrons on long distances even at elevated temperature could in principle be demonstrated using these (111) engineered QW~\cite{kikkawa_lateral_1999,couto_anisotropic_2007,hu_room_2011} . The control of the built-in piezoelectric field in strained (111) InGaAs quantum well could also permit to reduce significantly the DP spin relaxation even in the absence of an applied external field. This would open new perspectives for spin manipulation and control in solid state systems.\\
Acknowledgments : This work was supported by the France-China NSFC-ANR research project SPINMAN (grant 10911130356); we thank Prof. M.W. Wu for useful discussions and  S. Pinaud, G. Almuneau and A. Arnoult for technical support.


\begin{thebibliography}{41}
\expandafter\ifx\csname natexlab\endcsname\relax\def\natexlab#1{#1}\fi
\expandafter\ifx\csname bibnamefont\endcsname\relax
  \def\bibnamefont#1{#1}\fi
\expandafter\ifx\csname bibfnamefont\endcsname\relax
  \def\bibfnamefont#1{#1}\fi
\expandafter\ifx\csname citenamefont\endcsname\relax
  \def\citenamefont#1{#1}\fi
\expandafter\ifx\csname url\endcsname\relax
  \def\url#1{\texttt{#1}}\fi
\expandafter\ifx\csname urlprefix\endcsname\relax\def\urlprefix{URL }\fi
\providecommand{\bibinfo}[2]{#2}
\providecommand{\eprint}[2][]{\url{#2}}

\bibitem[{\citenamefont{Meier and Zakharchenya}(1984)}]{Meier1984}
\bibinfo{author}{\bibfnamefont{F.}~\bibnamefont{Meier}} \bibnamefont{and}
  \bibinfo{author}{\bibfnamefont{B.~P.} \bibnamefont{Zakharchenya}},
  \emph{\bibinfo{title}{{Optical Orientation}}} (\bibinfo{publisher}{Elsevier
  Science Ltd}, \bibinfo{year}{1984}).

\bibitem[{\citenamefont{Awschalom et~al.}(2002)\citenamefont{Awschalom, Loss,
  and Samarth}}]{awschalom_semiconductor_2010}
\bibinfo{author}{\bibfnamefont{D.}~\bibnamefont{Awschalom}},
  \bibinfo{author}{\bibfnamefont{D.}~\bibnamefont{Loss}}, \bibnamefont{and}
  \bibinfo{author}{\bibfnamefont{N.}~\bibnamefont{Samarth}},
  \emph{\bibinfo{title}{Semiconductor Spintronics and Quantum Computation}}
  (\bibinfo{publisher}{Springer}, \bibinfo{year}{2002}).

\bibitem[{\citenamefont{Dyakonov}(2008)}]{Dyakonov_book}
\bibinfo{author}{\bibfnamefont{M.}~\bibnamefont{Dyakonov}},
  \emph{\bibinfo{title}{Spin Physics in Semiconductors}}
  (\bibinfo{publisher}{Springer, New York}, \bibinfo{year}{2008}).

\bibitem[{\citenamefont{Dyakonov and Perel'}(1972)}]{Dyakonov1972}
\bibinfo{author}{\bibfnamefont{M.}~\bibnamefont{Dyakonov}} \bibnamefont{and}
  \bibinfo{author}{\bibfnamefont{V.}~\bibnamefont{Perel'}},
  \bibinfo{journal}{Soviet Physics Solid State} \textbf{\bibinfo{volume}{13}},
  \bibinfo{pages}{3023} (\bibinfo{year}{1972}).

\bibitem[{\citenamefont{Dyakonov and Kachorovskii}(1986)}]{Dyakonov1986}
\bibinfo{author}{\bibfnamefont{M.}~\bibnamefont{Dyakonov}} \bibnamefont{and}
  \bibinfo{author}{\bibfnamefont{V.}~\bibnamefont{Kachorovskii}},
  \bibinfo{journal}{Soviet Physics Solid State} \textbf{\bibinfo{volume}{20}},
  \bibinfo{pages}{110} (\bibinfo{year}{1986}).

\bibitem[{\citenamefont{\ifmmode \check{Z}\else
  \v{Z}\fi{}uti\ifmmode~\acute{c}\else \'{c}\fi{}
  et~al.}(2004)\citenamefont{\ifmmode \check{Z}\else
  \v{Z}\fi{}uti\ifmmode~\acute{c}\else \'{c}\fi{}, Fabian, and
  Das~Sarma}}]{zutic}
\bibinfo{author}{\bibfnamefont{I.}~\bibnamefont{\ifmmode \check{Z}\else
  \v{Z}\fi{}uti\ifmmode~\acute{c}\else \'{c}\fi{}}},
  \bibinfo{author}{\bibfnamefont{J.}~\bibnamefont{Fabian}}, \bibnamefont{and}
  \bibinfo{author}{\bibfnamefont{S.}~\bibnamefont{Das~Sarma}},
  \bibinfo{journal}{Rev. Mod. Phys.} \textbf{\bibinfo{volume}{76}},
  \bibinfo{pages}{323} (\bibinfo{year}{2004}).

\bibitem[{\citenamefont{Ivchenko}(2005)}]{ivchenko_optical_2005}
\bibinfo{author}{\bibfnamefont{E.~L.} \bibnamefont{Ivchenko}},
  \emph{\bibinfo{title}{Optical Spectroscopy of Semiconductor Nanostructures}}
  (\bibinfo{publisher}{Alpha Science International, Ltd},
  \bibinfo{year}{2005}).

\bibitem[{\citenamefont{Dresselhaus}(1955)}]{dresselhaus_spin-orbit_1955}
\bibinfo{author}{\bibfnamefont{G.}~\bibnamefont{Dresselhaus}},
  \bibinfo{journal}{Phys. Review} \textbf{\bibinfo{volume}{100}},
  \bibinfo{pages}{580} (\bibinfo{year}{1955}).

\bibitem[{\citenamefont{Bychkov and Rashba}(1984)}]{bychkov_oscillatory_1984}
\bibinfo{author}{\bibfnamefont{Y.~A.} \bibnamefont{Bychkov}} \bibnamefont{and}
  \bibinfo{author}{\bibfnamefont{E.~I.} \bibnamefont{Rashba}},
  \bibinfo{journal}{J. of Physics C} \textbf{\bibinfo{volume}{17}},
  \bibinfo{pages}{6039} (\bibinfo{year}{1984}).

\bibitem[{\citenamefont{Krebs and Voisin}(1996)}]{krebs_giant_1996}
\bibinfo{author}{\bibfnamefont{O.}~\bibnamefont{Krebs}} \bibnamefont{and}
  \bibinfo{author}{\bibfnamefont{P.}~\bibnamefont{Voisin}},
  \bibinfo{journal}{Phys. Rev. Lett.} \textbf{\bibinfo{volume}{77}},
  \bibinfo{pages}{1829} (\bibinfo{year}{1996}).

\bibitem[{\citenamefont{Malinowski et~al.}(2000)}]{malinowski_spin_2000}
\bibinfo{author}{\bibfnamefont{A.}~\bibnamefont{Malinowski}}
  \bibnamefont{et~al.}, \bibinfo{journal}{Phys. Rev. B}
  \textbf{\bibinfo{volume}{62}}, \bibinfo{pages}{13034} (\bibinfo{year}{2000}).

\bibitem[{\citenamefont{Amand et~al.}(1997)}]{amand_spin_1997}
\bibinfo{author}{\bibfnamefont{T.}~\bibnamefont{Amand}} \bibnamefont{et~al.},
  \bibinfo{journal}{Phys. Rev. Lett.} \textbf{\bibinfo{volume}{78}},
  \bibinfo{pages}{1355} (\bibinfo{year}{1997}).

\bibitem[{\citenamefont{Bu\ss{} et~al.}(2010)}]{bu_long_2010}
\bibinfo{author}{\bibfnamefont{J.~H.} \bibnamefont{Bu\ss{}}}
  \bibnamefont{et~al.}, \bibinfo{journal}{Appl. Phys. Lett.}
  \textbf{\bibinfo{volume}{97}}, \bibinfo{pages}{062101}
  (\bibinfo{year}{2010}).

\bibitem[{\citenamefont{Lagarde et~al.}(2008)}]{Lagarde2008}
\bibinfo{author}{\bibfnamefont{D.}~\bibnamefont{Lagarde}} \bibnamefont{et~al.},
  \bibinfo{journal}{Phys. Rev. B} \textbf{\bibinfo{volume}{77}},
  \bibinfo{pages}{041304} (\bibinfo{year}{2008}).

\bibitem[{\citenamefont{Averkiev and Golub}(1999)}]{averkiev_giant_1999}
\bibinfo{author}{\bibfnamefont{N.~S.} \bibnamefont{Averkiev}} \bibnamefont{and}
  \bibinfo{author}{\bibfnamefont{L.~E.} \bibnamefont{Golub}},
  \bibinfo{journal}{Phys. Rev. B} \textbf{\bibinfo{volume}{60}},
  \bibinfo{pages}{15582} (\bibinfo{year}{1999}).

\bibitem[{\citenamefont{Kainz et~al.}(2003)}]{kainz_anisotropic_2003}
\bibinfo{author}{\bibfnamefont{J.}~\bibnamefont{Kainz}} \bibnamefont{et~al.},
  \bibinfo{journal}{Phys. Rev. B} \textbf{\bibinfo{volume}{68}},
  \bibinfo{pages}{075322} (\bibinfo{year}{2003}).

\bibitem[{\citenamefont{Miller et~al.}(2003)}]{miller_gate-controlled_2003}
\bibinfo{author}{\bibfnamefont{J.~B.} \bibnamefont{Miller}}
  \bibnamefont{et~al.}, \bibinfo{journal}{Phys. Rev. Lett.}
  \textbf{\bibinfo{volume}{90}}, \bibinfo{pages}{076807}
  (\bibinfo{year}{2003}).

\bibitem[{\citenamefont{Eldridge et~al.}(2011)}]{eldridge_spin-orbit_2011}
\bibinfo{author}{\bibfnamefont{P.~S.} \bibnamefont{Eldridge}}
  \bibnamefont{et~al.}, \bibinfo{journal}{Phys. Rev. B}
  \textbf{\bibinfo{volume}{83}}, \bibinfo{pages}{041301}
  (\bibinfo{year}{2011}).

\bibitem[{\citenamefont{Jusserand et~al.}(1995)}]{jusserand_spin_1995}
\bibinfo{author}{\bibfnamefont{B.}~\bibnamefont{Jusserand}}
  \bibnamefont{et~al.}, \bibinfo{journal}{Phys. Rev. B}
  \textbf{\bibinfo{volume}{51}}, \bibinfo{pages}{4707} (\bibinfo{year}{1995}).

\bibitem[{\citenamefont{Schliemann
  et~al.}(2003)}]{schliemann_nonballistic_2003}
\bibinfo{author}{\bibfnamefont{J.}~\bibnamefont{Schliemann}}
  \bibnamefont{et~al.}, \bibinfo{journal}{Phys. Rev. Lett.}
  \textbf{\bibinfo{volume}{90}}, \bibinfo{pages}{146801}
  (\bibinfo{year}{2003}).

\bibitem[{\citenamefont{Lechner et~al.}(2009)}]{lechner_tuning_2009}
\bibinfo{author}{\bibfnamefont{V.}~\bibnamefont{Lechner}} \bibnamefont{et~al.},
  \bibinfo{journal}{Appl. Phys. Lett.} \textbf{\bibinfo{volume}{94}},
  \bibinfo{pages}{242109} (\bibinfo{year}{2009}).

\bibitem[{\citenamefont{Stich et~al.}(2007)}]{stich_detection_2007}
\bibinfo{author}{\bibfnamefont{D.}~\bibnamefont{Stich}} \bibnamefont{et~al.},
  \bibinfo{journal}{Phys. Rev. B} \textbf{\bibinfo{volume}{76}},
  \bibinfo{pages}{073309} (\bibinfo{year}{2007}).

\bibitem[{\citenamefont{Averkiev et~al.}(2006)}]{averkiev_spin-relaxation_2006}
\bibinfo{author}{\bibfnamefont{N.~S.} \bibnamefont{Averkiev}}
  \bibnamefont{et~al.}, \bibinfo{journal}{Phys. Rev. B}
  \textbf{\bibinfo{volume}{74}}, \bibinfo{pages}{033305}
  (\bibinfo{year}{2006}).

\bibitem[{\citenamefont{Liu et~al.}(2007)}]{liu_electron_2007}
\bibinfo{author}{\bibfnamefont{B.}~\bibnamefont{Liu}} \bibnamefont{et~al.},
  \bibinfo{journal}{Appl. Phys. Lett.} \textbf{\bibinfo{volume}{90}},
  \bibinfo{pages}{112111} (\bibinfo{year}{2007}).

\bibitem[{\citenamefont{Koralek et~al.}(2009)}]{koralek_emergence_2009}
\bibinfo{author}{\bibfnamefont{J.~D.} \bibnamefont{Koralek}}
  \bibnamefont{et~al.}, \bibinfo{journal}{Nature}
  \textbf{\bibinfo{volume}{458}}, \bibinfo{pages}{610} (\bibinfo{year}{2009}).

\bibitem[{\citenamefont{Ohno et~al.}(1999)}]{ohno_spin_1999}
\bibinfo{author}{\bibfnamefont{Y.}~\bibnamefont{Ohno}} \bibnamefont{et~al.},
  \bibinfo{journal}{Phys. Rev. Lett.} \textbf{\bibinfo{volume}{83}},
  \bibinfo{pages}{4196} (\bibinfo{year}{1999}).

\bibitem[{\citenamefont{D\"ohrmann et~al.}(2004)}]{doehrmann_anomalous_2004}
\bibinfo{author}{\bibfnamefont{S.}~\bibnamefont{D\"ohrmann}}
  \bibnamefont{et~al.}, \bibinfo{journal}{Phys. Rev. Lett.}
  \textbf{\bibinfo{volume}{93}}, \bibinfo{pages}{147405}
  (\bibinfo{year}{2004}).

\bibitem[{\citenamefont{Lau and Flatt\'e}(2002)}]{lau_tunability_2002}
\bibinfo{author}{\bibfnamefont{W.~H.} \bibnamefont{Lau}} \bibnamefont{and}
  \bibinfo{author}{\bibfnamefont{M.~E.} \bibnamefont{Flatt\'e}},
  \bibinfo{journal}{J. Appl. Phys.} \textbf{\bibinfo{volume}{91}},
  \bibinfo{pages}{8682} (\bibinfo{year}{2002}).

\bibitem[{\citenamefont{Karimov et~al.}(2003)}]{karimov_high_2003}
\bibinfo{author}{\bibfnamefont{O.~Z.} \bibnamefont{Karimov}}
  \bibnamefont{et~al.}, \bibinfo{journal}{Phys. Rev. Lett.}
  \textbf{\bibinfo{volume}{91}}, \bibinfo{pages}{246601}
  (\bibinfo{year}{2003}).

\bibitem[{\citenamefont{Cartoix\`a et~al.}(2005)}]{cartoixa_suppression_2005}
\bibinfo{author}{\bibfnamefont{X.}~\bibnamefont{Cartoix\`a}}
  \bibnamefont{et~al.}, \bibinfo{journal}{Phys. Rev. B}
  \textbf{\bibinfo{volume}{71}}, \bibinfo{pages}{045313}
  (\bibinfo{year}{2005}).

\bibitem[{\citenamefont{Vurgaftman and
  Meyer}(2005)}]{vurgaftman_spin-relaxation_2005}
\bibinfo{author}{\bibfnamefont{I.}~\bibnamefont{Vurgaftman}} \bibnamefont{and}
  \bibinfo{author}{\bibfnamefont{J.~R.} \bibnamefont{Meyer}},
  \bibinfo{journal}{J. Appl. Phys.} \textbf{\bibinfo{volume}{97}},
  \bibinfo{pages}{053707} (\bibinfo{year}{2005}).

\bibitem[{\citenamefont{Sun et~al.}(2010)}]{sun_spin_2010}
\bibinfo{author}{\bibfnamefont{B.~Y.} \bibnamefont{Sun}} \bibnamefont{et~al.},
  \bibinfo{journal}{J. Appl. Phys.} \textbf{\bibinfo{volume}{108}},
  \bibinfo{pages}{093709} (\bibinfo{year}{2010}).

\bibitem[{\citenamefont{Tober and Bahder}(1993)}]{tober_determining_1993}
\bibinfo{author}{\bibfnamefont{R.~L.} \bibnamefont{Tober}} \bibnamefont{and}
  \bibinfo{author}{\bibfnamefont{T.~B.} \bibnamefont{Bahder}},
  \bibinfo{journal}{Appl. Phys. Lett.} \textbf{\bibinfo{volume}{63}},
  \bibinfo{pages}{2369} (\bibinfo{year}{1993}).

\bibitem[{\citenamefont{Vinattieri et~al.}(1993)}]{vinattieri_electric_1993}
\bibinfo{author}{\bibfnamefont{A.}~\bibnamefont{Vinattieri}}
  \bibnamefont{et~al.}, \bibinfo{journal}{Appl. Phys. Lett.}
  \textbf{\bibinfo{volume}{63}}, \bibinfo{pages}{3164} (\bibinfo{year}{1993}).

\bibitem[{\citenamefont{Leyland et~al.}(2007)}]{leyland_oscillatory_2007}
\bibinfo{author}{\bibfnamefont{W.~J.~H.} \bibnamefont{Leyland}}
  \bibnamefont{et~al.}, \bibinfo{journal}{Phys. Rev. B}
  \textbf{\bibinfo{volume}{76}}, \bibinfo{pages}{195305}
  (\bibinfo{year}{2007}).

\bibitem[{\citenamefont{Eldridge et~al.}(2008)}]{eldridge_all-optical_2008}
\bibinfo{author}{\bibfnamefont{P.~S.} \bibnamefont{Eldridge}}
  \bibnamefont{et~al.}, \bibinfo{journal}{Phys. Rev. B}
  \textbf{\bibinfo{volume}{77}}, \bibinfo{pages}{125344}
  (\bibinfo{year}{2008}).

\bibitem[{not()}]{note}
\bibinfo{note}{We observe a negligible variation of the $g$ factor when $E$
  increases. A weak $\Delta g/g$ has been measured (not shown) but it plays here a
  negligible role in the damping of the oscillations.}

\bibitem[{\citenamefont{Bernardes et~al.}(2007)}]{bernardes_spin-orbit_2007}
\bibinfo{author}{\bibfnamefont{E.}~\bibnamefont{Bernardes}}
  \bibnamefont{et~al.}, \bibinfo{journal}{Phys. Rev. Lett.}
  \textbf{\bibinfo{volume}{99}}, \bibinfo{pages}{076603}
  (\bibinfo{year}{2007}).

\bibitem[{\citenamefont{Kikkawa and Awschalom}(1999)}]{kikkawa_lateral_1999}
\bibinfo{author}{\bibfnamefont{J.~M.} \bibnamefont{Kikkawa}} \bibnamefont{and}
  \bibinfo{author}{\bibfnamefont{D.~D.} \bibnamefont{Awschalom}},
  \bibinfo{journal}{Nature} \textbf{\bibinfo{volume}{397}},
  \bibinfo{pages}{139} (\bibinfo{year}{1999}).

\bibitem[{\citenamefont{Couto et~al.}(2007)}]{couto_anisotropic_2007}
\bibinfo{author}{\bibfnamefont{O.~D.~D.} \bibnamefont{Couto}}
  \bibnamefont{et~al.}, \bibinfo{journal}{Phys. Rev. Lett.}
  \textbf{\bibinfo{volume}{98}}, \bibinfo{pages}{036603}
  (\bibinfo{year}{2007}).

\bibitem[{\citenamefont{Hu et~al.}(2011)}]{hu_room_2011}
\bibinfo{author}{\bibfnamefont{C.}~\bibnamefont{Hu}} \bibnamefont{et~al.},
  \bibinfo{journal}{Nanoscale Research Letters} \textbf{\bibinfo{volume}{6}},
  \bibinfo{pages}{149} (\bibinfo{year}{2011}).

\end{thebibliography}
\end{document}